# Ultraviolet dielectric hyperlens with layered graphene and boron nitride

Junxia Wang[1,2], Yang Xu[1, 4,a)], Hongsheng Chen[1,2,4, b)], and Baile Zhang[3, c)]

[1]Department of Information Science and Electronic Engineering, Zhejiang University, Hangzhou 310027, China

[2]The Electromagnetics Academy at Zhejiang University, Zhejiang University, Hangzhou 310027, China

[3]Division of Physics and Applied Physics, School of Physical and Mathematical Sciences, Nanyang Technological University, Singapore 637371

[4]Cyrus Tang Center for Sensor Materials and Applications, Zhejiang University, Hangzhou 310027, China

**Abstract**: The concept of hyperlens, as a novel transformation optics device, is a promising real-time super-resolution lens that can effectively transform evanescent waves into propagating waves and thus break the diffraction limit. However, previous hyperlens implementations usually adopted metal which would absorb most energy during light transmission and thus deteriorate imaging efficiency. Here we propose a novel hyperlens design based on dielectric layered graphene and $h$-boron nitride ($h$-BN) whose performance can surpass the counterpart design with metal. Our first-principle and Kramers-Kronig relation calculation shows that both layered graphene and layered $h$-BN exhibit strong anisotropic properties in ultraviolet spectrum regions, where their permittivity components perpendicular to the optic axis can be negative while the components parallel to the optic axis can be positive. Based on the anisotropic properties, flat and cylindrical hyperlenses are designed and numerically verified with layered graphene at 1200 THz and layered $h$-BN at 1400 THz, respectively. Our work provides a dielectric hyperlens approach to overcome the diffraction limit at ultraviolet frequencies, which may find applications where dynamic imaging of subwavelength features at the molecular and cellular scales is desired.

**Corresponding Authors:**

a) *yangxu-isee@zju.edu.cn*

b) *hansomchen@zju.edu.cn*

c) *blzhang@ntu.edu.sg*

Super-resolution, that can break the fundamental physical law of "diffraction limit" discovered in 1873 [1], has brought many breakthroughs in the past few decades in biology, medicine, material science, and chemistry. As the key to understanding more fundamental dynamical process at the molecular and cellular scales, real-time super-resolution, that is able to capture subwavelength dynamical information, will provide a powerful tool for next revolution in related disciplines. One approach to real-time super-resolution is the "superlens" proposed in Ref. [2], which utilized negative-refractive-index metamaterials to recover evanescent fields [3,4]. However, this method provided no magnification and could not deliver the subwavelength information to the far field [5]. In view of this flaw, researchers have proposed the concept of "hyperlens" [6,7,8,9], which converted evanescent waves to propagating ones and provided magnification. The hyperlens is named after its hyperbolic dispersion of strongly anisotropic metamaterials that can be implemented by stacking multi-layers of alternating metal and dielectrics. However, metal is well-known lossy at optical frequencies, and should be avoided in some applications such as fluorescence imaging whose signal usually is very week. It is thus strongly desirable to have a dielectric design of hyperlens that can reduce the loss during light transmission in practice.

Graphene, a recently discovered two-dimensional (2D) hexagonal crystal carbon sheet [10], has unique optoelectronic properties [11,12] with great potential to solve bottlenecks in many electronic/photonic technologies. For example, Ref. [13] utilized graphene to design novel transformation optics devices, where superlens and Luneburg lens are numerically demonstrated. This ingenious application apparently is a big step toward real breakthrough of graphene optics. However, different from the superlens and Luneburg lens possessing isotropic optical properties, a hyperlens with anisotropic optical properties has not been discussed with possible implementation of graphene. Here we fill this gap by designing a hyperlens based on layered graphene and layered $h$-boron nitride. Our design provides a unique subwavelength real-time imaging approach at ultraviolet frequencies that can potentially find applications in photonics, material science, and chemistry.

To implement a hyperlens, we need to utilize anisotropy. Let us first examine the structure of graphene sheets. Because of the hexagonal crystal structure, carbon atoms in the basal plane are bonded together by strong covalent bonds while the adjacent sheets are tightly stacked through Van der waals forces [14,15]. The unique structures give rise to significant anisotropy [16], which is ideal for hyperlens implementation. The dielectric constant tensor of graphene has two independent components: $\varepsilon_\perp = \varepsilon_{R\perp} + i\varepsilon_{i\perp}$, $\varepsilon_{//} = \varepsilon_{R//} + i\varepsilon_{i//}$, where $\varepsilon_\perp$ and $\varepsilon_{//}$ are the permittivity components perpendicular and parallel to the optic axis, respectively. M. Klintenberg et al. [17] found the real part of graphene $\varepsilon_\perp$ is negative between 4.3 and 6.5 eV in the optical spectrum. Because of the 2D hexagonal similarity to graphene, monolayer $h$-boron nitride ($h$-BN) [18,19] also exhibits anisotropic optical properties, which stimulus our interest to apply them on hyperlens.

Integration of individual 2D material sheets into macroscopic structures without breaking their properties is essential for the application of graphene and $h$-boron nitride. Recent graphene layered structures including folding of graphene sheets by mechanical deformation [20] and 10-20 nm graphene films by CVD growth [21] are reported. These 3D composites of graphene not only preserve electronic and optical properties of 2D graphene but also show good flexibility, namely, they can be bended, stretched, and twisted without breaking. Therefore, these dielectric anisotropic composites can be useful to design electromagnetic devices including hyperlens.

We first study the anisotropic permittivity tensor of monolayer graphene and *h*-boron nitride with the density functional theory in the generalized gradient approximation (GGA) using the optical package of SIESTA [22]. The double $\zeta$ polarized numerical atomic basis sets and norm-conserving pseudo-potentials are used. Already-standard corrections like GGA+U are mainly used to calculate redox reaction energies of many transition metal compounds, since the self-interaction error in LDA and GGA is not canceled out in redox reactions where an electron is transferred between significantly different environments [23-25]. Because we only calculate the optical properties of graphene and *h*-boron nitride without any metal oxides, GGA is chosen here. The exchange-correlation function of GGA is represented by the Revised Perdew- Burke- Ernzerh (RPBE) approximation [26]. A Kgrid-Monkhorst-Pack mesh of $250 \times 250 \times 1$ (monolayer graphene and *h*-BN) and a 300 Ry energy cutoff are used to ensure converged GGA results. To investigate the anisotropy properties, different polarized light is applied to the graphene and *h*-BN plane with Gaussian broadening of 0.02 eV. In order to fulfill the *f*-sum rule, it is necessary to set the energy range 0-40 eV or even wider. An optical mesh of $800 \times 800 \times 4$ is used, which determines the mesh size used for the integration across the Brillouin zone related to the calculation accuracy.

The anisotropic dielectric constant tensors of the two structures are obtained as shown in Figure 1, which are consistent with previous work in Refs. [17,27,28]. Using these anisotropic dielectric constants, we design flat and cylindrical hyperlenses at ultraviolet frequencies with graphene and

*h*-boron nitride layers. Although experimental measurements on $\varepsilon_{//}$ for monolayer graphene are rarely investigated, our simulated dielectric constants between 0.83 eV and 9.12 eV (200 THz-2200 THz), as shown in Fig. 1(a), agree well with Ref. [17]. More detailed comparison between the results in Ref. [17] and our results can be seen in Supplementary Material. It is worth mentioning that with increase of the number of graphene layers, the dielectric constants do not change much in the ultraviolet range, especially above 1200 THz. In Fig. 1(a), an intense resonance near 4.2 eV (1011 THz) is found due to the intra-layer transitions among π bands for $\varepsilon_{\perp}$. In the range of 4.2 eV-6.8 eV (1011 THz-1645 THz), the real part of $\varepsilon_{\perp}$ is negative. In contrast, for $\varepsilon_{//}$, no resonance below 10 eV is found because the 2D selection rules forbid π→π* transitions (<10 eV) while allowing σ→σ* instead [29]. The dielectric properties of monolayer *h*-BN are shown in Fig. 1(b). For $\varepsilon_{\perp}$ component, we see the resonance frequency of monolayer *h*-BN occurs near 5.6 eV (1342 THz), higher than that of graphene. The real part of $\varepsilon_{\perp}$ is negative in the range of 5.6 eV-7.52 eV (1342 THz-1815 THz), where $\varepsilon_{//}$ is almost constant. From the theoretical calculation, we see that both graphene and *h*-BN possess strong anisotropy properties due to their 2D atomic structures, which is very useful to implement hyperlens.

We then briefly introduce the mechanism of hyperlens that is originally discussed in Ref. [6,8] in a simplified flat slab model. Consider a slab with anisotropic permittivity Diag($\varepsilon_x, \varepsilon_y, \varepsilon_z$) shown in Fig. 2(a) where a plane wave with transverse-magnetic (TM) polarization is incident on the slab. The dispersion relation in isotropic medium such as free space in Region 0 is represented by $k_x^2 + k_z^2 = \omega^2 \mu_0 \varepsilon_0$, where $k_x$ and $k_z$ are wave vectors along *x* and *z* axis, $\varepsilon_0$ and $\mu_0$ are permittivity and permeability in free space respectively, and $\omega$ is the angular frequency. It is well known that the accessible range of $k_x$ in the imaging plane determines the imaging resolution. In conventional isotropic medium, the dispersion wave number *k*-surface is a circle, as shown in Fig. 2(b). When $k_x$ is larger than the wave number in free space $k_0$, $k_z$ is an imaginary number, meaning that the wave decays exponentially when propagating along the z direction and thus becomes evanescent. Therefore the subwavelength information, which is carried by the waves with $k_x > k_0$, decays very fast and cannot reach the far field. In anisotropic medium with different signs of permittivity components, the dispersion relation is a hyperbolic curve represented by $k_x^2/\varepsilon_z + k_z^2/\varepsilon_x = \omega^2 \mu_y$ ($\varepsilon_x > 0$ and $\varepsilon_z < 0$ in Fig. 2(b)). The derivation of the dispersion curve is shown in Supplementary Material. For waves with arbitrarily large $k_x$, $k_z$ always has a real number solution. Therefore the finer subwavelength information can be preserved when propagating in the medium. To achieve better performance, the dispersion curve should be almost flat such that the image and the object only differ by a phase because $k_z$ is almost constant.

Here we provide a new perspective to understand hyperlens by using transformation optics, as shown in Fig. 3. Top panels of Fig. 3 are *k* surfaces for materials with different dispersion relations, while the bottom are the corresponding change in virtual space from transformation optics point of view. The first column is for free space where the light energy will spread out during propagation because of diffraction, as can be seen in Fig. 3(d). To overcome the energy spreading associated with diffraction, one possibility is to squeeze the free space heavily in the horizontal direction. Figure 3(e) shows that the space becomes like a very thin tube and all energy will go through the space along this tube without spreading out. Therefore the diffraction limit can be overcome. As a result, according to the principle of transformation optics [30,31], $\varepsilon_z$ will go to infinity in the limit and the dispersion curve becomes a very flat ellipse. It should be emphasized that the essence of this space squeezing is to create a flat dispersion curve, while the shape, whether it is elliptical or hyperbolic, is a secondary factor. Although this space squeezing can give intuitive explanation on the flatness of dispersion curve, in practice a very large $\varepsilon_z$ is almost impractical to obtain. To facilitate fabrication while maintaining the flatness of dispersion curve, a compromised way is to switch to the negative value for $\varepsilon_z$, which is the case in Fig. 3(c) and (f). Although the dispersion is thus changed from parabolic to hyperbolic, the central part of *k* surface is still very flat. Light still propagates along a thin tube without spreading out as shown in Fig. 3(f), therefore choosing an appropriate negative value for $\varepsilon_z$ and creating a hyperbolic dispersion curve is a more practical way, while the mechanism is the same as for the flat elliptical dispersion curve. Based on the analysis above, the quantitative calculation and simulation are given to show the flat hyperlens effect. Suppose a TM plane wave with $\overline{k_i} = \hat{x} k_{ix} + \hat{z} k_{iz}$ is incident from Region 0 as in Fig. 2(a). The transmission coefficient T from Region 0 to Region 2 can be expressed as

$$T = \frac{\overline{H_2}(z=d)}{\overline{H_0}(z=0)} = \frac{4 e^{i(k_{1z}-k_{2z})d}}{(1+p_{01})\cdot(1+p_{12})+(1-p_{01})\cdot(1-p_{12})e^{i2k_{1z}d}}, \quad (1)$$

where

$$p_{01} = \frac{\varepsilon_0 k_{1z}}{\varepsilon_z k_{iz}}, \quad p_{12} = \frac{\varepsilon_z k_{2z}}{\varepsilon_0 k_{1z}}.$$

The detailed calculation process is in the Supplemental Material. The transmission coefficient T is plotted as a function of the transverse wave vector $k_x$ in Fig. 4. The thickness of slab is 70 nm. In the case of free space in Region 1, T=1 for the propagating waves ($k_x < k_0$), and it decays exponentially for the evanescent waves ($k_x > k_0$). However, for layered graphene based hyperlens with the parameter $\varepsilon_x = 2.229 + 0i$ and $\varepsilon_z = -3.817 + 4.265i$ at 1200 THz, there is also significant transmission even for large wave vectors and it drops much slower with increase of $k_x$. The case of silver in Region 1 ($\varepsilon = -0.213 + 3.639i$ [32]) shows that at this frequency silver based flat superlens does not provide any recovery of subwavelength information, although it worked previously at lower frequencies [3]. The main reason is because at this frequency, the permittivity of silver has a much larger imaginary part compared to its real part, which means the light energy will be easily dissipated in propagation.

To better illustrate how layered graphene and layered *h*-boron nitride can be used to achieve hyperlenses, we simulate two concrete imaging configurations in flat and cylindrical structures in COMSOL MULTIPHYSICS by adopting the electromagnetic parameters obtained from first-principle and Kramers-Kronig relation calculation. At ultraviolet frequency band, materials are non-magnetic, *i.e.* $\mu_z = \mu_0$, which is in agreement with that described in Ref. [33]. As illustrated in Fig. 1(a), $\varepsilon_\perp$ is negative in the frequency region from 1011 THz to 1645 THz for graphene while $\varepsilon_{//}$ is positive. We choose the frequency of 1200 THz where $\varepsilon_\perp = -3.815 + 4.265i$ and $\varepsilon_{//} = 2.229 + 0i$, as we have mentioned previously. Simulation result in Fig. 5(a) shows the imaging of a flat hyperlens with thickness of 70 nm, based on layered graphene with its carbon plane aligned perpendicular to *x*-axis. It is also possible to implement this hyperlens by simply cutting from a bulk of graphite with thickness of 70 nm. Two point sources are located along *x*-axis and separated by 70 nm which is about a quarter of the wavelength in vacuum at 1200 THz. Due to the hyperbolic dispersion in Region 1, the evanescent waves from Region 0 can be converted to propagating ones in Region 1. Figure 5(b) shows the $H_y$ field distribution when the inner rectangular slab is air. The comparison of the intensity in the imaging plane for air and layered graphene based hyperlens are shown respectively in Figure 5 (d), from which we see that in the imaging plane, the two sources can still be distinguished after propagating through a hyperlens while they cannot be distinguished after propagating through air. It should be emphasized that although $\varepsilon_z$ has an imaginary part comparable to that of metal, the final transmission still has little loss. This is because the light is mainly propagating parallel to *z*-axis, and the loss introduced by the imaginary part of $\varepsilon_z$ is thus very limited. To better demonstrate the superiority of this hyperlens, we compare its performance with previous silver/$Al_2O_3$ based hyperlens at the frequency of 1200 THz. Due to the dispersion characteristics of metal, dielectric tensor of silver comes to be $\varepsilon = -0.213 + 3.639i$ at 1200 THz as mentioned in Ref. [32] and the permittivity of $Al_2O_3$ at this frequency is $\varepsilon = 3.24 + 0i$ [34]. Figure 5(c) shows the $H_y$ field distribution when the slab is a stack of silver/$Al_2O_3$ and the intensity on the imaging plane is also given in Fig. 5(d) for comparison. The simulation results tell that silver/$Al_2O_3$ based hyperlens fails to provide super-resolution at this high frequency.

Now we have understood the mechanism of hyperlens and have designed a flat hyperlens based on layered graphene. However, the flat slab geometry cannot provide magnification, which, on the other hand, is very important in some super-resolution applications. Let us proceed to create magnification by changing geometry from flat to cylindrical. Graphene layers as well as *h*-boron nitride can be integrated into Chinese fan structures while maintaining the optical properties of 2D monolayer [21]. Therefore, the strong anisotropy characteristics are also preserved.

The schematic of a typical cylindrical hyperlens is shown in Fig. 6(a), where graphene or *h*-BN layers are arranged into Chinese fan region. The graphene/*h*-BN plane is aligned with radial directions ($\rho$). The reason why this cylindrical structure can provide magnification can be intuitively explained in Fig. 6(b-e) by transformation optics theory. The virtual space of a hyperlens in Fig. 6(e) (k-surface in Fig. 6(b)) can be thought of as the original space in Fig. 6(d) (k-surface in Fig. 6(c)) heavily squeezed in the azimuthal direction. The light energy propagates along this squeezed thin tube without spreading out. Apparently, when the light energy of the two point sources propagates outwards, the distance between them will be magnified. Therefore, a magnified image is formed on the imaging plane.

We then simulate this Chinese fan hyperlens in COMSOL MULTIPHYSICS by adopting the optical parameters of layered graphene at 1200 THz and *h*-boron nitride at 1400 THz. Figure 7(a) presents $H_y$ field distribution of two sources at the inner semicircle of the hyperlens with 200 nm thickness. The distance between the two sources is 100 nm which is less than half of the wavelength in vacuum. (The previously demonstrated hyperlens achieved 130 nm resolutions [9].) Regarding the boundary conditions of the simulation region in COMSOL MULTIPHYSICS, we set the out boundaries as scattering boundary conditions and all the inner boundaries as continuous boundary conditions. Because monolayer graphene can be folded into sub-micro-scale thickness which can be realized by roll-to-roll process or CVD growth [21,35], we vary the thickness of hyperlens from 200 nm to 400 nm while keeping the distance between sources constant. For the graphene hyperlens with 400 nm thickness, the simulated $H_y$ field distribution is shown in Fig. 7(c). Figure 7(e) presents $H_y$ field distribution of two sources without hyperlens. (Also, we have given the simulation results of silver based superlens and hyperlens in Supplementary Material at the same frequency for comparison. It shows that at 1200 THz, silver based superlens as well as hyperlens cannot overcome the diffract limitation.) It is clear to see that two sources can be distinguished in the far field with hyperlens while in air they cannot. Figure 7(g) compares the intensity in the imaging plane for layered graphene hyperlens and air dielectric. For the air case, only one peak exists in the imaging plane. While for the graphene hyperlens case, two peaks exist, indicating that subwavelength resolution is achieved by the graphene-based hyperlens.

For *h*-BN layers, the negative dielectric constants are located between 1342 THz and 1815 THz.

Thus we choose 1400 THz as the operational frequency of h-BN with $\varepsilon_\perp = -1.637 + 1.839i$ and $\varepsilon_{//} = 2.277 + 0i$. The simulated hyperlens results based on the BN are shown in Fig. 7 (b, d, f, h). The distance between two sources for h-BN hyperlens is 100 nm approximately equal to $1/2\lambda$. Figure 7(b) and 7(d) show the $H_y$ field distribution for h-BN hyperlens with 200 nm and 400 nm thickness, respectively. Figure 7(h) compares the intensity in the imagine plane for h-BN hyperlens and air dielectric. From Fig. 7 we see the resolution of h-BN based hyperlens is more sensitive to the thickness than graphene layers.

Another interesting question is whether we can use thin graphite directly to implement the hyperlens. We have calculated the dielectric tensors for different polarized waves from one to six layers of graphene and thin graphite (The results are shown in Supplementary Material). The optical couplings between carbon layers are mainly located in low-energy frequency range. For our choice of 1200 THz for graphene and 1400 THz for h-boron nitride, the influence of coupling are relatively weak. Figure 8 shows the simulation results of thin-layer graphite compared to the layered graphene. The parameters of monolayer graphene we obtained from *ab initio* calculations are $\varepsilon_\rho = -3.815 + 4.265i$ and $\varepsilon_\varphi = 2.229 + 0i$, while the corresponding parameters of thin-layer graphite are $\varepsilon_\rho = -3.817 + 4.086i$ and $\varepsilon_\varphi = 2.131 + 0.213i$. From Fig. 8 we can see that the thin graphite and multilayer graphene have similar hyperlens effect.

In conclusion, we studied the strong anisotropy properties of layered graphene and layered h-boron nitride from the first-principle and Kramers-Kronig relation calculation, and explored their potential applications in hyperlens. Our work shows that the permittivity components perpendicular to the optic axis can be negative in ultraviolet region and the one parallel to the optic axis can be positive, which is desirable for hyperlens implementation. Flat and cylindrical hyperlenses are subsequently designed with super-resolution performance beyond the previous silver based superlens and silver/$Al_2O_3$ based hyperlens at 1200 THz. A *h*-boron nitride hyperlens is similarly designed at 1400 THz. Our work provides a purely dielectric approach to hyperlens design and is thus of significant practical value to real-time super-resolution imaging applications.

Acknowledgements: This work is sponsored by the NNSFC (Grants Nos. 60990320, 60990322, 60977043 and 61006077), the NBRPC (Grant No. 2007CB613405), State Key Program of NSFC (Grant No. 10834004), the FANEDD (Grant No. 200950), the ZJNSF (Grant No. R1080320, R12F040001), the SRFDP with Grant Nos. 20100101120045, and the NTU SUG. The authors thank Shanghai Supercomputer Center for simulation support and Innovation plat form for Micro/Nano Device and System Integration at Zhejiang University.

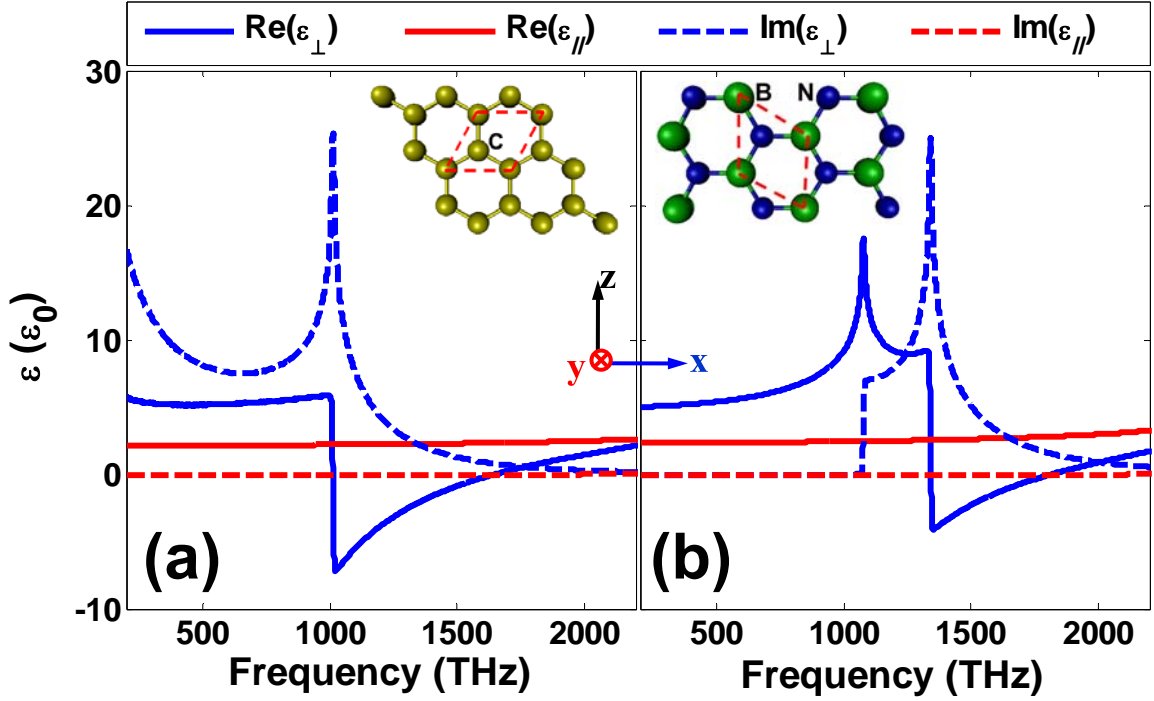

Fig. 1: The anisotropy permittivity tensor for (a) graphene (GR) and (b) *h*-boron nitride (*h*-BN) ( "⊥" represents *x* and *z* components and "//" represents the *y* component). The insets show the atom unit cell of graphene and BN for dielectric constant calculation. (a) $\varepsilon_\perp$ of graphene shows a resonance at 4.3 eV (1000 THz) and the real part of $\varepsilon_\perp$ is negative in the range of 4.2 eV-6.8 eV (1011 THz-1645 THz). $\varepsilon_{//}$ remains a positive constant during 4.2 eV-6.8 eV. (b) $\varepsilon_\perp$ of *h*-BN shows a resonance at ~5.6 eV (~1350 THz), which is higher than that of graphene. The real part of $\varepsilon_\perp$ is negative between 5.6 eV and 7.52 eV (1342-1815 THz) where $\varepsilon_{//}$ is positive.

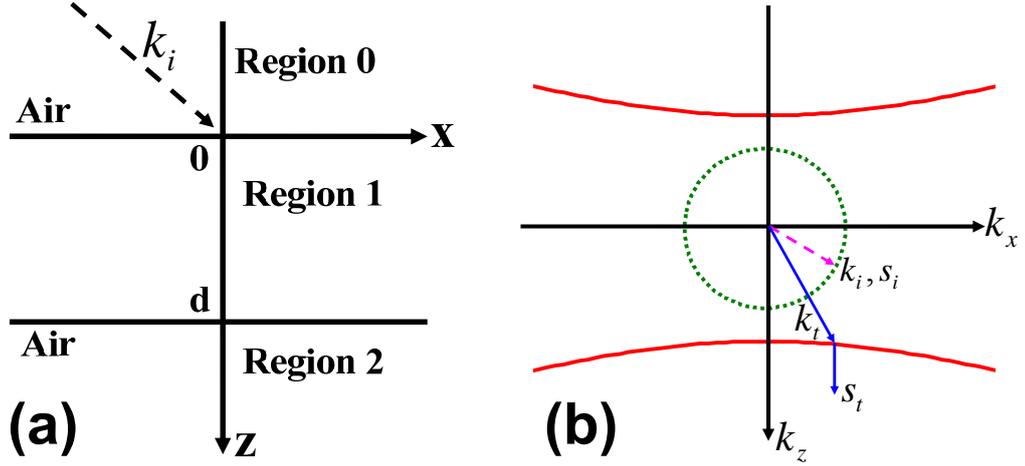

Fig. 2: (a) Configuration of a plane wave incident upon a slab with thickness *d*; (b) Green circle: dispersion curve of light in free space. Red curve: hyperbolic dispersion in an anisotropic medium where $\varepsilon_x > 0$ and $\varepsilon_z < 0$.

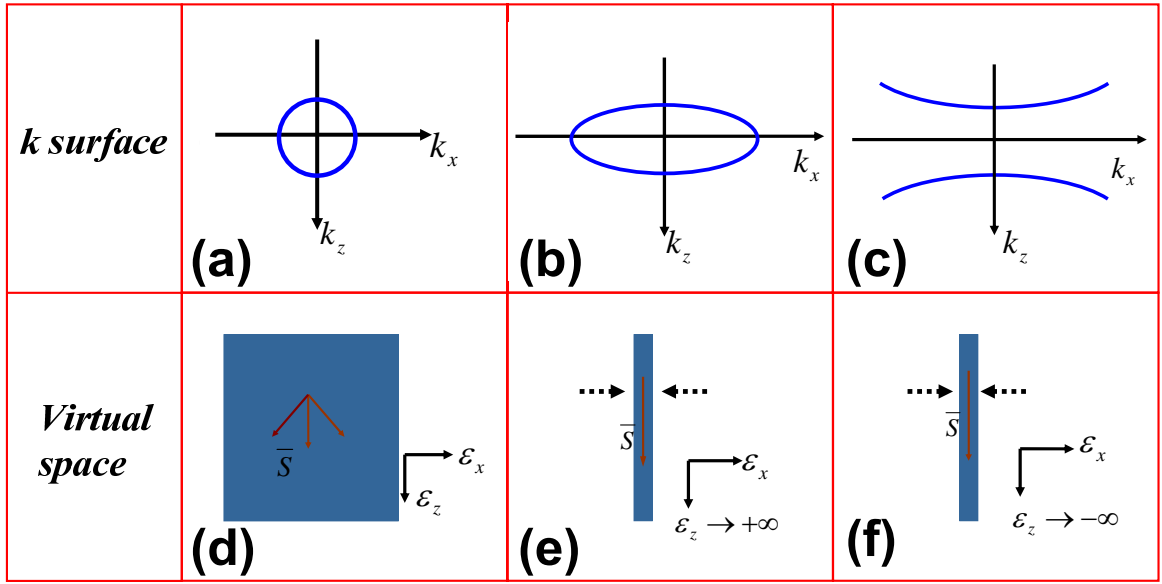

Fig. 3: Transform schematic of flat hyperlens. Top panel (a,b,c) are *k* surfaces for different materials while bottom (d,e,f) are corresponding virtual spaces. (a) and (d) are for free space in which energy will spread out. (b) and (e) correspond to the case of $\varepsilon_z$ approaching infinity, when free space is squeezed heavily in the horizontal direction and energy will not spread out. (c) and (f) are the case when $\varepsilon_z$ comes to be negative infinity. Because the central part of *k* surface is the same as (b), energy flow in such a material still has good directivity.

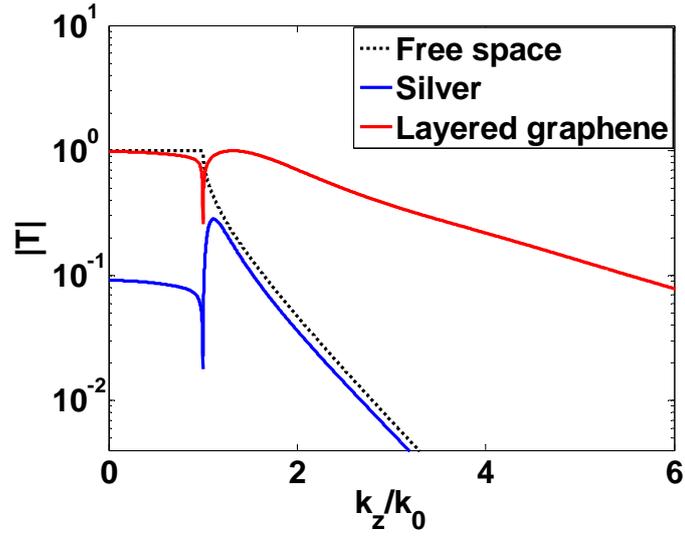

Fig. 4: Calculated results of T vs. $k_x$ for different materials in Region 1 with the thickness of 70 nm. Red: layered graphene in Region 1 with $\varepsilon_z = -3.817 + 4.265i$ and $\varepsilon_x = 2.229 + 0i$ at 1200 THz; Blue: silver with $\varepsilon = -0.213 + 3.639i$ at the same frequency; Black: free space in Region 1.

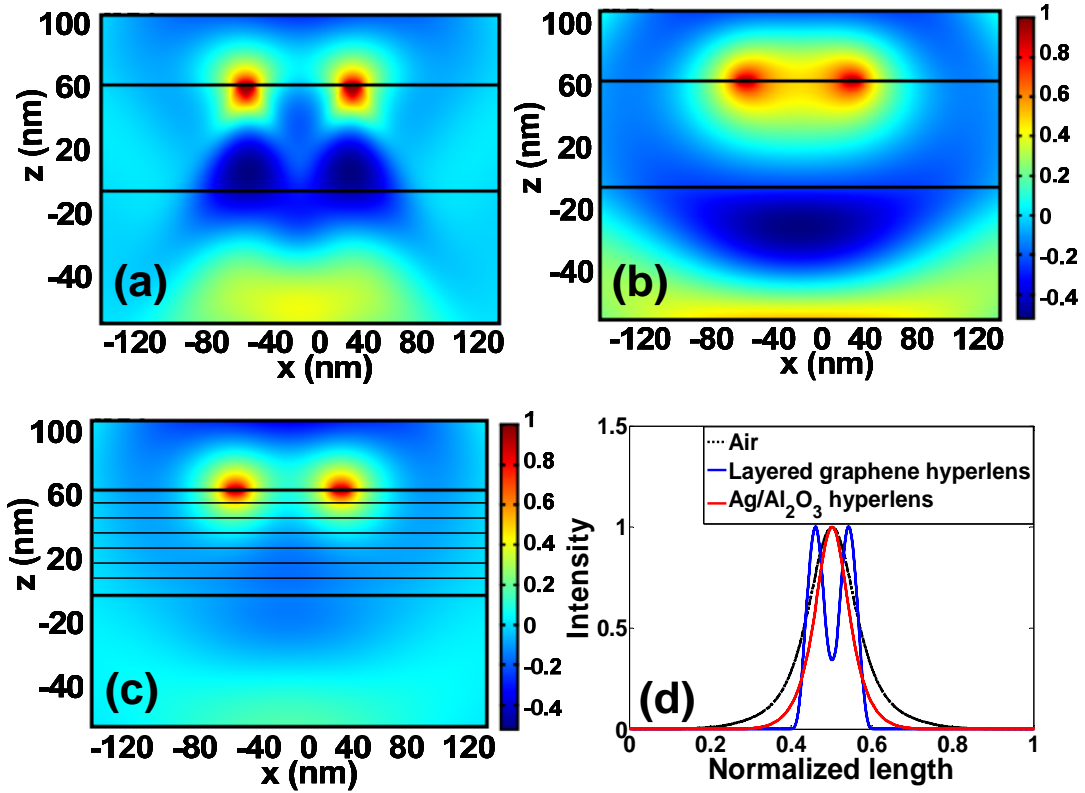

Fig. 5 (a) $H_y$ field distribution with layered graphene in the inner rectangular region with $\varepsilon_z = -3.817 + 4.265i$ and $\varepsilon_x = 2.229 + 0i$. The inner rectangular region's thickness is 70 nm and two sources located along *x*-axis with the distance of 70 nm. (b) $H_y$ field distribution with air in the inner rectangular region. (c) $H_y$ field distribution with a stack of silver and $Al_2O_3$ layers at 1200 THz. (d) The comparison of intensities in the image plane for air, layered graphene, and silver/$Al_2O_3$, respectively.

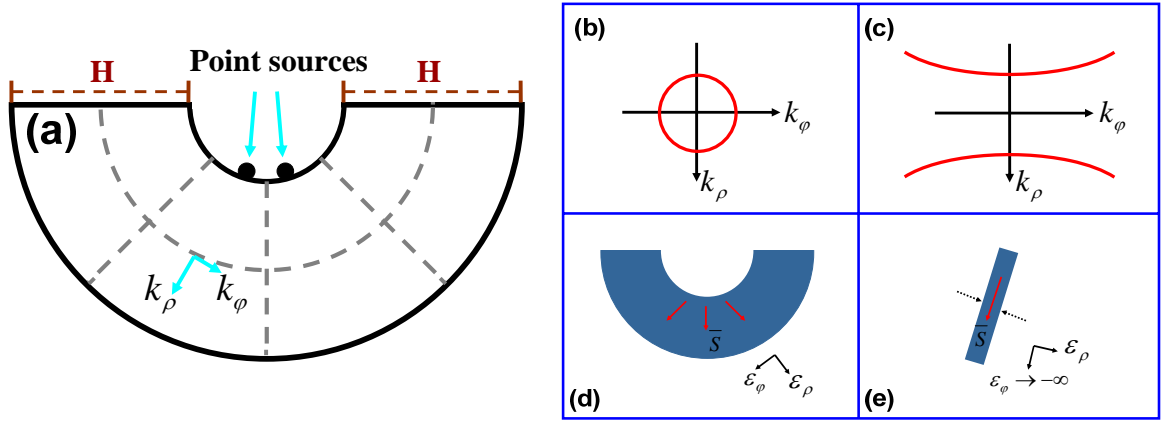

Fig. 6: (a) The schematic of a cylindrical hyperlens with two point sources closely located in the inner semicircle. Each fan-shaped region is fully filled with graphene or *h*-BN structure layers with graphene/*h*-BN plane aligned with axial directions ($\rho$). H is the thickness of hyperlens based on 2D layered materials. (b) and (c) are *k* surfaces for different materials while bottom (d) and (e) are corresponding virtual spaces. (b) and (d) are isotropic material in which energy will spread out. (c) and (e) indicate when $\varepsilon_z$ becomes negative infinity, the original virtual space is heavily squeezed in the azimuthal direction and energy will not spread out.

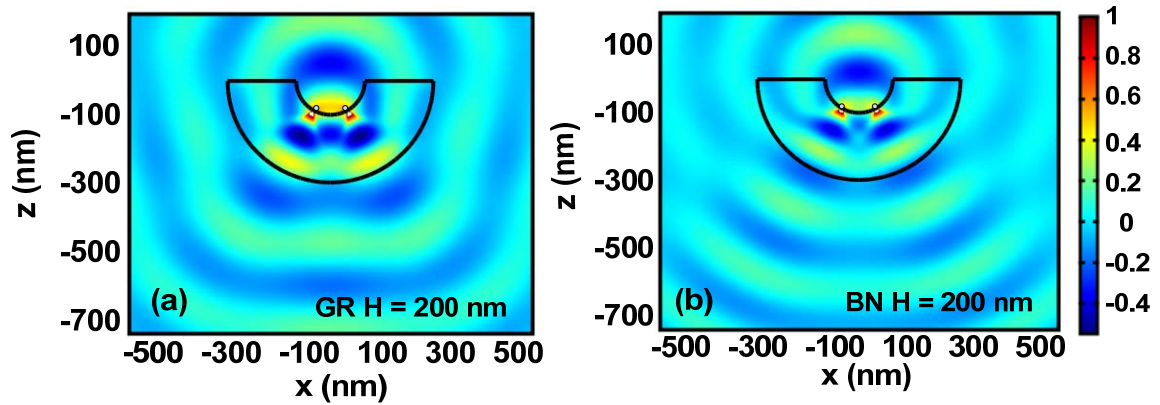

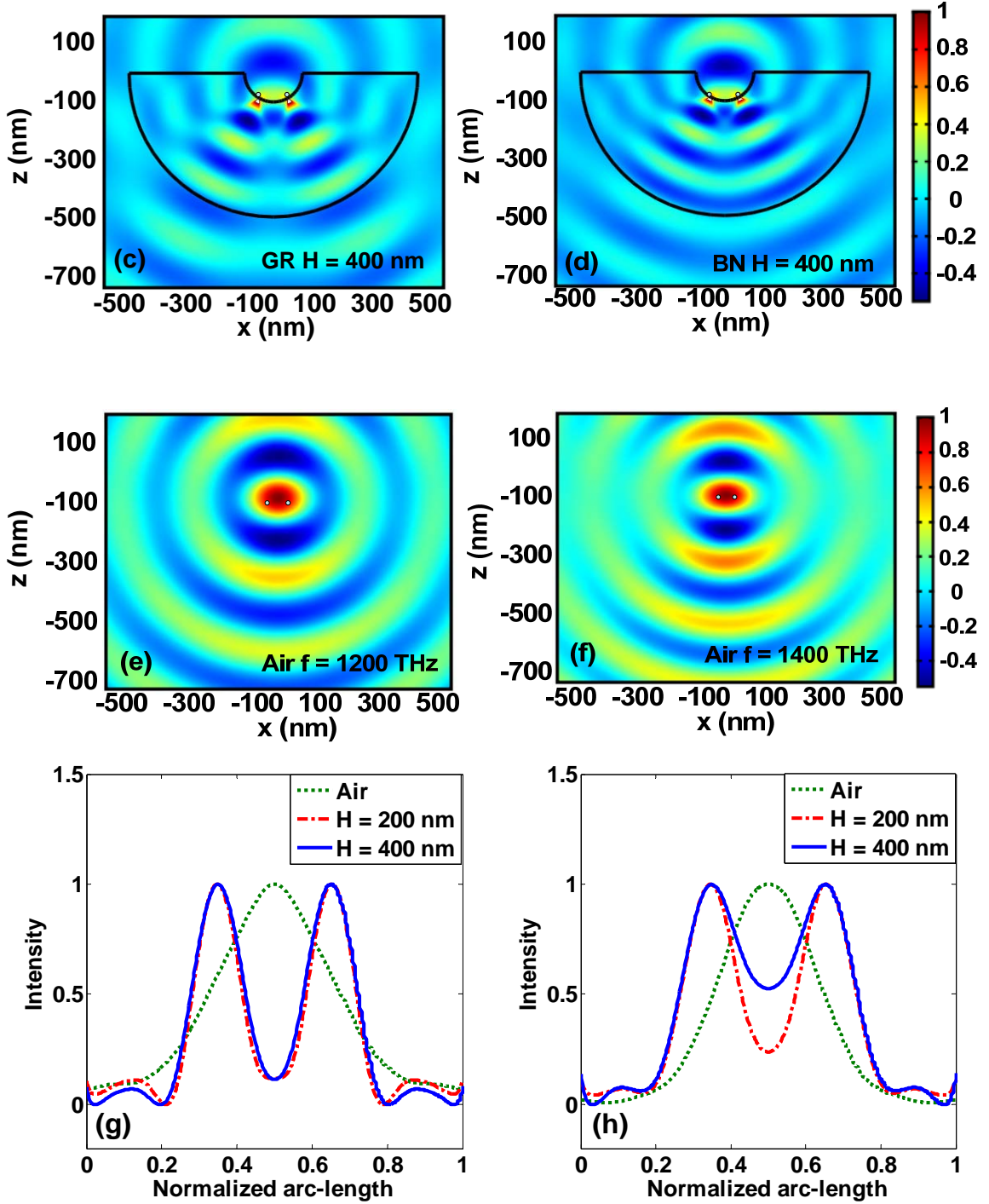

Fig. 7: Simulation results of hyperlens based on layered graphene (GR) and layered *h*-boron nitride (*h*-BN). Graphene at frequency of 1200 THz is shown in (a,c), with $\varepsilon_\rho = -3.815 + 4.265i$ and $\varepsilon_\varphi = 2.229 + 0i$. The distance between two sources is 100 nm, which is less than 1/2 vacuum wavelength. *h*-BN layers at 1400 THz are shown in (b,d) with $\varepsilon_\rho = -1.637 + 1.839i$ and $\varepsilon_\varphi = 2.277 + 0i$. The distance between two sources is the same as graphene. $H_y$ field distribution of two sources propagating through a hyperlens varying from 200 nm to 400 nm are shown in (a,c) for graphene and in (b,d) for $h$-BN, respectively. (e) and (f) present $H_y$ field distribution of two sources propagating without hyperlens for comparison. Figure 6(g) and (h) compare the intensity in the imaging plane for graphene, $h$-BN, and air dielectric, respectively.

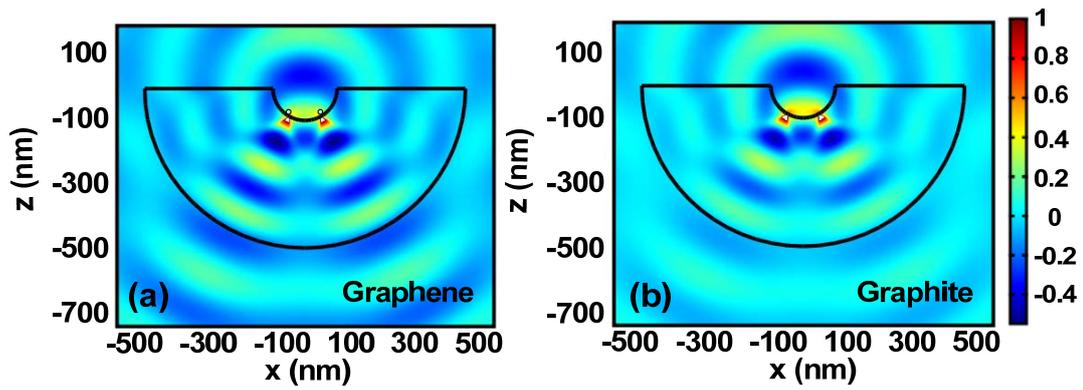

Figure 8: Simulation results by using parameters of (a) layered graphene at 1200 THz and (b) thin-layer graphite at the same frequency.